\documentclass[12pt,preprint]{aastex}

\shorttitle{Cluster of Radio Sources in NGC~2024}
\shortauthors{Rodr\'\i guez, G\'omez, \& Reipurth}

\begin{document}

\title{A Cluster of Compact Radio Sources in NGC~2024 (Orion B)}

\author{Luis F. Rodr\'\i guez and Yolanda G\'omez} 
\affil{Centro de Radioastronom\'\i a y Astrof\'\i sica, UNAM, 
Apdo. Postal 3-72 (Xangari), 58089 Morelia, Michoac\'an, M\'exico}
\email{l.rodriguez, y.gomez@astrosmo.unam.mx}

\and

\author{Bo Reipurth}
\affil{Institute for Astronomy, University of Hawaii, 2680 Woodlawn Drive, Honolulu, HI 96822}
\email{reipurth@ifa.hawaii.edu}

\begin{abstract}

We present deep 3.6~cm radio continuum observations
of the H~II region NGC~2024 in Orion B obtained
using the Very Large Array in its A-configuration,
with $0\rlap.{''}2$ angular resolution. We detect a total
of 25 compact radio sources in a region of $4' \times 4'$.
We discuss the nature of these sources and its relation with the
infrared and X-ray objects in the region.
At least two of the radio sources are obscured proplyds whose morphology
can be used to restrict the location of the main ionizing
source of the region. This cluster
of radio sources is compared with others that have been 
found in regions of recent star formation.

\end{abstract}  


\section{INTRODUCTION}

NGC~2024 is a prominent H~II region located in the Orion B (L 1630)
star-forming complex, at a distance of $\sim$415 pc
(Anthony-Twarog 1982). The H~II region is sharply ionization-bounded
to the south (Barnes et al. 1989). With a total flux density of
$\sim$60 Jy at cm wavelengths (Rodr\'\i guez \& Chaisson 1978;
Barnes et al. 1989), the H~II region requires about 
$10^{48}$ ionizing photons per second to maintain
its ionization, a flux that can be provided by an O9 ZAMS star
with a luminosity of $\sim 5.0 \times 10^4~L_\odot$.  
Until recently, the required exciting star had not been identified,
but Bik et al. (2003) have proposed that the infrared source IRS2b
is an embedded late O or early B star responsible for the ionization.

A prominent dust lane is seen in optical images to
run across the region from north to south.
This dust lane is evident also in molecular line
observations of the region (i. e. Schulz et al. 1991).
Mezger et al. (1988) identified 6 small-scale submm condensations 
embedded along this dense ridge,
suggesting that star formation is presently taking place here.

Clusters of near-infrared (Lada et al. 1991; Beck et al. 2003; Haisch et al. 2000), mid-infrared 
(Haisch et al. 2001), and X-ray sources (Skinner, Belzer, \&
Gagner 2003) are known to exist in association with NGC~2024.
Furthermore, a number of submillimeter continuum sources were found
and studied by Mezger et al. (1988) and Visser et al. (1998). To provide a more
complete census of the young stellar population in this
region, we have undertaken deep, high angular resolution 3.6 cm observations in an attempt
to detect a cluster of compact radio sources similar to those
found in Orion A (Garay, Moran, \& Reid 1987;
Churchwell et al. 1987), NGC~1333 (Rodr\'\i guez, Anglada, \& Curiel 1999),
the Arches region near the galactic center (Lang, Goss, \& Rodr\'\i guez 2001),
and in GGD~14 (G\'omez, Rodr\'\i guez, \& Garay 2000; 2002). 

\section{OBSERVATIONS}

We have used the Very Large Array of the NRAO\footnote{The National Radio Astronomy
Observatory is a facility of the National Science Foundation operated
under cooperative agreement by Associated Universities, Inc.} 
in its A-array configuration to observe NGC~2024
at 3.6~cm. The region was observed on
2002 March 2, 3, and 8. Our phase center was at
$\alpha(J2000) = 05^h 41^m 44\rlap.{^s}9;~ \delta(J2000) = -01^\circ 55' 54{''}$.
The amplitude calibrator was
1331+305, with an adopted flux density of 5.18 Jy,
and the phase calibrator was 0541$-$056, with an average
flux density of 0.748$\pm$0.004 Jy over the
three epochs of observation. The data were analyzed in the standard manner
using the package AIPS of NRAO. The data were self-calibrated in phase. 
Individual images were made at each epoch to search for fast variability
(on a timescale of days)
between the epochs observed.
To diminish the presence of extended emission, in particular that originating from the
very sharp ionization front of the H~II region that is
evident in Fig. 1b of Barnes et al. (1989), we used only visibilities
with baselines larger than 100 k$\lambda$, thus suppressing the emission
of structures larger than $\sim$2${''}$.
The images were restored with a circular beam of $0\rlap.{''}24$,
the average value of the angular resolution of the individual maps
made with the ROBUST parameter of IMAGR set to 0.
The three maps were then averaged to obtain an rms noise of 15 $\mu$Jy.

A total of 25 sources were detected in a region of $4' \times 4'$.
The distribution of these sources is shown in Figure 1.
Following Fomalont et al. (2002), we estimate that in a field
of $4' \times 4'$ the \sl a priori \rm number of expected 3.6 cm sources 
above 0.1 mJy is $\sim$0.6. We then conclude that probably
one out of the 25 sources could be a background object, but that we are
justified in assuming that practically all the members of the radio cluster are 
associated with NGC~2024.
In Figure 2 we show the positions of the radio sources overlapped on the 
red DSS2 image of the region. From this figure it is evident that the
cluster of radio sources is closely associated with the central parts
of the dust lane that runs
across NGC~2024.

In Table 1 we list the positions and flux densities of the sources,
averaged over the three observations. We also note
in column 5
if they were found to be variable or not. The number
given in parenthesis for the variable sources is the ratio between the largest and
smallest flux density observed. Finally, in the
last three columns we list counterparts, when found. A counterpart was taken
as such if its position was within 3$''$ of the radio position.
The radio positions are estimated to be accurate to $\sim 0\rlap.{''}05$.

\section{OVERALL CHARACTERISTICS OF THE SAMPLE OF RADIO SOURCES}

Out of the 25 radio sources detected, 
13 have a 2MASS counterpart
and 15 have a Chandra counterpart (Skinner et al. 2003).
Only the brightest source in the radio cluster, VLA~19, had been previously
reported at radio wavelengths (Snell \& Bally 1986;
Gaume, Johnston, \& Wilson
1992; Kurtz, Churchwell, \& Wood 1994).
Only 4 of the 25 radio sources, VLA~8, VLA~12, VLA~17,
and VLA~18, lack a previously reported counterpart.

Of the 25 sources detected, 8 were found to be time variable over
the observed timescale of a few days (see column 5 in Table 1).
We searched for linear and circular polarization in the sample and
only one source (VLA 24) showed left circular polarization
at the 3\% level. 
We measured the angular size of the sources using the AIPS
task IMFIT, with a correction for bandwidth smearing.
With the exception of sources 
VLA~8, VLA~13, and VLA~19, all sources were found to be 
unresolved, $\theta_s \leq 0\rlap.{''}2$.

We found radio continuum sources closely associated with three of the
six submm sources of Mezger et al. (1988), namely FIR 4 (VLA~9),
FIR 5 (VLA~10) and FIR 6
(VLA~14). All three sources are relatively weak and do not
show time variability. Remarkably, none of these three radio sources
have a near-IR or X-ray counterpart, suggesting very large
extinction toward them.

Two close ($\leq 2''$) double systems, formed by sources VLA~12 and 13,
as well as VLA~15 and 16, respectively, are part of the cluster.
Most probably they constitute physical binaries.

\section{COMMENTS ON SELECTED INDIVIDUAL SOURCES}

\subsection{VLA~8}

This source has no reported counterparts.
It is one of the few sources that is clearly resolved,
elongated in the north-south
direction (see Fig. 3).
Its deconvolved dimensions are 
$0\rlap.{''}59 \pm 0\rlap.{''}03 \times 0\rlap.{''}25 \pm 0\rlap.{''}02;
PA = 177^\circ \pm 3^\circ$.
These angular dimensions correspond to 170 AU $\times$ 95 AU.
One possible explanation is that we are observing a thermal radio
jet (e. g. Rodr\'\i guez 1997) that could be powering the 
unipolar redshifted CO jet in the region (Richer, Hills, \& Padman 1992),
since both the radio source and the CO jet are well aligned and elongated
in the north-south
direction. However, the CO jet seems to emanate from a point about
1$'$ south of VLA~8, a position much closer to VLA~10 (=FIR~5).
Furthermore, the detailed shape of the source is curved,
unlike most thermal jets that are rather straight.
This curved morphology is reminiscent of that seen in
cometary H~II regions or in ionized proplyds.
It is known
that the ionized Orion proplyds, when observed in the radio continuum 
(Henney et al. 2002), show an arc-shaped structure.
Furthermore, the physical dimensions of VLA~8 are similar to those
of the Orion proplyd LV~2 (Henney et al. 2002).  
Clearly, additional 
observations are needed to establish if
this radio source is a thermal jet, a cometary H~II region, or a proplyd.
As we will see below, the fact that the arc
``points'' to the region where the ionizing star of the region is believed to
be located
favors the identification of this source as a radio proplyd.

\subsection{VLA~9}

This source has no reported near-IR or X-ray counterpart.
It is associated with FIR~4, one of the small-scale submm condensations
reported by Mezger et al. (1988), and is thus likely to represent a deeply
embedded protostar. It is associated with an infrared reflection
nebula (Moore \& Yamashita 1995) and a unipolar redshifted CO outflow
(Chandler \& Carlstrom 1996).

\subsection{VLA~10}

This source has no reported near-IR or X-ray counterpart,
but it is associated with FIR~5 (Mezger et al. 1988). It could be the powering source
of the unipolar redshifted CO jet studied by Richer et al. (1992)
and Chandler \& Carlstrom (1996)
since it is located at the position from where this jet seems to originate.
Wiesemeyer et al. (1997) present interferometric 3 mm
continuum observations, and show that FIR 5 separates into two clumps,
FIR 5e and 5w. VLA 10 coincides precisely with FIR 5w. Lai et
al. (2002) studied the detailed magnetic field structure around FIR 5.

\subsection{VLA~12 and VLA~13}

These two sources (see Fig. 4) form a close binary system
separated by $0\rlap.{''}9$,
which at 415 pc corresponds to 375 AU in
projection.
Source 12 has no reported counterpart.
The source VLA~13 is angularly resolved, with deconvolved dimensions
of $0\rlap.{''}46 \pm 0\rlap.{''}03 \times 0\rlap.{''}12 \pm 0\rlap.{''}04;
PA = 110^\circ \pm 3^\circ$.
Its arc-shaped morphology and the fact that it
points to the region where the ionizing star of the region is believed to
be located
favors, as in the case of VLA~8, the identification of VLA~13 as a radio proplyd.

\subsection{VLA~14} 

This source has no reported near-IR or X-ray counterpart.
It is associated with FIR~6 (Mezger et al. 1988) and could be the powering source
of a compact bipolar CO outflow found in this region (Richer 1990;
Chandler \& Carlstrom 1996)
since the outflow emanates from a position coincident with VLA~14.

\subsection{VLA~15 and VLA~16}

Both these VLA objects are 2MASS and Chandra sources.
They form a close binary system
(see Fig. 5) 
separated by $1\rlap.{''}7$. 
The source VLA~15 coincides positionally
with the infrared source IRS2b, that has been proposed to be the ionizing source of the
NGC~2024 H~II region by Bik et al. (2003).
The source IRS2b, first found by Jiang et al. (1984) and Nisini et al. (1994), 
is located about 5$''$ north-west of IRS2.
IRS2 is the brightest near-IR source of this region (detected by us as VLA~19, see below).
The source VLA~16 is not detected in the near-IR, but it is a Chandra source
(Skinner et al. 2003).

\subsection{VLA~17}

This source has no reported counterparts.
It is located exactly over the sharp ionization front that runs east-west
(Barnes et al. 1989). It could be
a bright knot in the ionization front,
with no young star directly associated.

\subsection{VLA~19}

This is the brightest radio source in the cluster, with near-IR and X-ray
counterparts. It was reported
previously as a radio source by Snell \& Bally (1986),
Gaume et al. (1992), and Kurtz et al. (1994). We found no significant variability
at the three epochs observed by us. However, our 3.6 cm flux density of 17.4 mJy
is about twice the value of 8.9 mJy reported by Kurtz et al. (1994), 
from VLA observations taken 13 years before. This 
radio source coincides with the infrared source IRS2 (Grasdalen 1974;
Barnes et al. 1989).
The radio source (Fig. 6) clearly shows structure. There is a bright, dominant
component to the west, with a fainter extension to the east.
At present it is not possible to establish if we are observing a binary source,
an ultracompact H~II region, or a radio proplyd. 
Multifrequency observations at
high angular resolution are needed to favor one of these possibilities.

\subsection{VLA~21} 

This source is quite bright in the radio and exhibits large variation
with a factor of 5.7.
It has a Chandra counterpart.

\subsection{VLA~24}

This source, with 2MASS and Chandra counterparts,
is the only one that showed evidence of polarization,
showing left circular polarization
at the 3\% level at all three epochs observed. 
It is also time variable with variations of about a factor of 2.
The combination of fast time variability
with circular polarization is an indicator of gyrosynchrotron
emission (Feigelson \& Montmerle 1999). 

\section{DISCUSSION}

There are several mechanisms that can produce compact centimeter radio
sources in regions of star formation. In regions of low mass star formation,
thermal jets and gyrosynchrotron emitters can be present.
In regions of high mass star formation we also have strong
ionizing radiation available and, in addition to the two
mechanisms present in low mass star-forming regions, we can have
ionized stellar winds, ultracompact H~II regions, and radio proplyds.
It is possible to distinguish between these various possibilities
with high angular resolution, multifrequency observations
made with the required sensitivity.
For NGC~2024 we only have available the 3.6 cm observations
presented here. We can, however, argue that because of its time variability and
circular polarization, VLA~24 is most probably a young low mass star
showing gyrosynchrotron emission.
In the case of sources VLA~2, 5, 6, 11, 20, 21, and 23, their 
fast time variability suggests also a
gyrosynchrotron nature.

Sources VLA~8 and VLA~13 are probably radio proplyds given their morphology and
orientation. If so, they can be used in an
attempt to search for the ionizing
source of the region. These two sources are also affected by bandwidth
smearing, but the deconvolution made correcting for this effect 
clearly shows that they are truly extended.
In Figure 7 we show an image that includes these two
sources as well as other nearby sources. 
In this image we have passed two lines by each of the
sources. These lines have position angles
of $\pm$3 $\sigma$
with respect to the perpendicular to the major axes of the sources.
The set of lines defines a four-sided region that includes 
VLA~15 (= IRS2b) and VLA~19 (=IRS2), as can be seen in Fig. 7. 
Given the uncertainties of the method, we cannot favor either of the sources
conclusively. We note, however, that the morphology of VLA~19 (=IRS2)
is suggestive also of a proplyd nature (see discussion in section 4.8 as well
as Fig. 6). If this is the case, then the object VLA~15 (= IRS2b) is clearly favored
as the ionizing source of the region, in agreement with Bik et al. (2003). 

Over the years, a few examples of radio clusters associated with
regions of recent star formation have been appearing in the literature
(Garay et al. 1987; Churchwell et al. 1987; Becker \& White 1988;
Stine \& O'Neal 1998; Rodr\'\i guez et al. 1999;
G\'omez et al. 2000; Lang et al. 2001).
In Table 2 we summarize the parameters of these radio clusters.
Although more such clusters should be studied to have a reliable statistical
base, some interesting trends are evident. The diameters of the radio clusters are in the
0.2 to 0.7 pc range. 
The radio luminosity of the most luminous member is correlated with
the bolometric luminosity of the cluster.
We believe that sensitive, high angular resolution studies similar to
that presented here are needed to establish if these clusters are
always present in other relatively nearby regions of massive star formation.
If properly understood, these radio clusters could be an invaluable observational
tool to study the stellar population of heavily obscured regions of star formation.

\section{CONCLUSIONS}

We have presented sensitive, high angular resolution ($0\rlap.{''}2$) VLA
observations at 3.6 cm toward the NGC 2024 region of
recent star formation.
Our main conclusions are summarized below.

1. We detected a total
of 25 compact radio sources in a region of $4' \times 4'$.
Only four of these sources do not have a
previously reported counterpart at any wavelength.
However, only one of the radio sources had been reported in the literature.

2. We found radio continuum sources closely associated with three of the
six submm sources in the region. Remarkably, none of these three radio/submm sources
has a near-IR or X-ray counterpart, suggesting very large
extinction toward them.

3. Two of the sources (VLA~8 and VLA~13) are proposed to be radio proplyds
whose study may help
pinpoint the origin of the ionizing radiation in the region.
Our attempts to do this suggest a region containing both VLA~15 (= IRS2b) and VLA~19 (=IRS2)
as the area where the ionizing source is located.

4. The source VLA~19, the counterpart of IRS2, is found to exhibit spatial structure, although
its precise nature remains undetermined. 
We also detected a radio counterpart to IRS2b, the source that has been
recently proposed to be ionizing NGC~2024.

5. Eight of the sources detected (VLA~2, 5, 6, 11, 20, 21, 23, and 24)
show fast time variability
and probably are young low mass stars exhibiting gyrosynchrotron
emission.

6. Two close ($\leq 2''$) binary systems, formed by sources VLA~12 and 13
and VLA~15 and 16, respectively, are part of the cluster.

7. We have briefly discussed the parameters of other radio clusters found in regions
of star formation and suggest that sensitive, high angular resolution
studies of other relatively nearby massive star formation regions are 
likely to detect similar clusters.

\vspace{0.5cm}

\acknowledgments

We thank Will Henney for his comments on 
proplyds and Steve Skinner for providing us with his list of
X-ray sources before publication.
We are also grateful to George Herbig for calling our attention
to the radio cluster in NGC~1579.
LFR and YG acknowledge the support
of DGAPA, UNAM, and of CONACyT (M\'exico).
BR acknowledges support from NASA grant NAG5-8108.
The Second Palomar Observatory Sky Survey (POSS-II) was 
made by the California Institute of Technology with funds from the
National Science Foundation, the National Aeronautics 
and Space Administration, the National Geographic Society, the Sloan
Foundation, the Samuel Oschin Foundation, 
and the Eastman Kodak Corporation. The Oschin Schmidt 
Telescope is operated by the
California Institute of Technology and Palomar Observatory. 
The 2MASS project is a collaboration between The University 
of Massachusetts and the Infrared Processing and Analysis 
Center (JPL/ Caltech), with funding provided primarily by NASA 
and the NSF. 
This research has made use of the SIMBAD database, 
operated at CDS, Strasbourg, France.

\clearpage

\figcaption{VLA continuum image of NGC~2024 at 3.6 cm.
The contour shown is 6
times 15 $\mu$Jy beam$^{-1}$, the rms noise of the image.
The half power full width of the circular restoring beam 
is $0\rlap.{''}24$. The compact sources are numbered from VLA~1 to
VLA~25. The faint emission features that appear to the east and
west of source VLA~17 are part of the ionization front and are not considered
individual compact sources.
\label{fig1}}

\figcaption{Position of the compact radio sources detected,
overlapped on a greyscale image of the region from the red Digital Sky Survey.
The position and number of some of the sources are shown in white so that
are seen better against the dark background.
\label{fig2}}

\figcaption{VLA continuum image of NGC~2024 VLA~8 at 3.6 cm.
The contours are $-$4, 4, 5, 6, 8, 10, 12, 15, and 20 
times 15 $\mu$Jy beam$^{-1}$, the rms noise of the image.
The half power contour of the circular restoring beam
is shown in the bottom left corner.
\label{fig3}}

\figcaption{VLA continuum image of the sources
NGC~2024 VLA~12 and 13 at 3.6 cm.
The contours are $-$4, 4, 5, 6, 8, 10, and 12
times 15 $\mu$Jy beam$^{-1}$, the rms noise of the image.
The half power contour of the circular restoring beam
is shown in the bottom left corner.
\label{fig4}}

\figcaption{VLA continuum image of the sources
NGC~2024 VLA~15 and 16 at 3.6 cm.
The contours are $-$4, 4, 5, 6, 8, 10, 12, 15, 20, and 30
times 15 $\mu$Jy beam$^{-1}$, the rms noise of the image.
The half power contour of the circular restoring beam
is shown in the bottom left corner.
The north-south elongation of the sources is due to the bandwidth smearing
aberration.
\label{fig5}}

\figcaption{VLA continuum image of the source
NGC~2024 VLA~19 at 3.6 cm.
The contours are $-$5, 5, 6, 8, 10, 12, 15, 20, 30,
40, 60, 80, 100, 150, 200, 300, and 400
times 15 $\mu$Jy beam$^{-1}$, the rms noise of the image.
The half power contour of the circular restoring beam
is shown in the bottom left corner.
The north-south elongation of the source is due to the bandwidth smearing
aberration.
\label{fig6}}

\figcaption{VLA continuum image of several sources in
the core of
NGC~2024 at 3.6 cm.
The contour shown is 6
times 15 $\mu$Jy beam$^{-1}$, the rms noise of the image.
The dashed lines have position angles
of $\pm$3 $\sigma$
with respect to the perpendicular to the major axes of the sources
VLA~8 and VLA~13.
These lines define a four-sided region that includes sources 
VLA~15 (IRS2b) and VLA~19 (IRS2).
Note that the ten VLA sources in this Figure are distributed in
the plane of the sky approximately in an ellipse.
\label{fig7}}

\newpage
\plotone{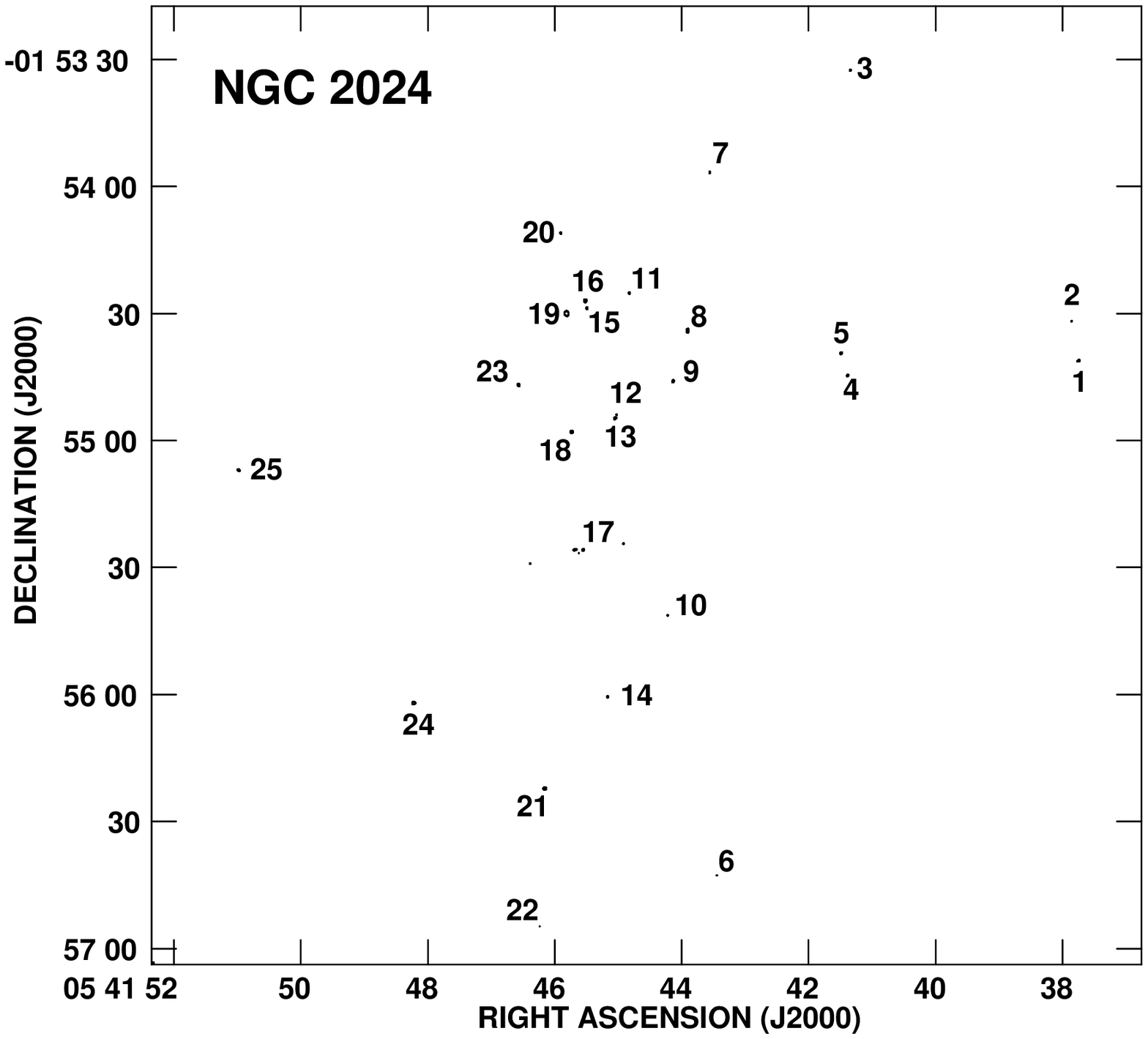}
\newpage
\plotone{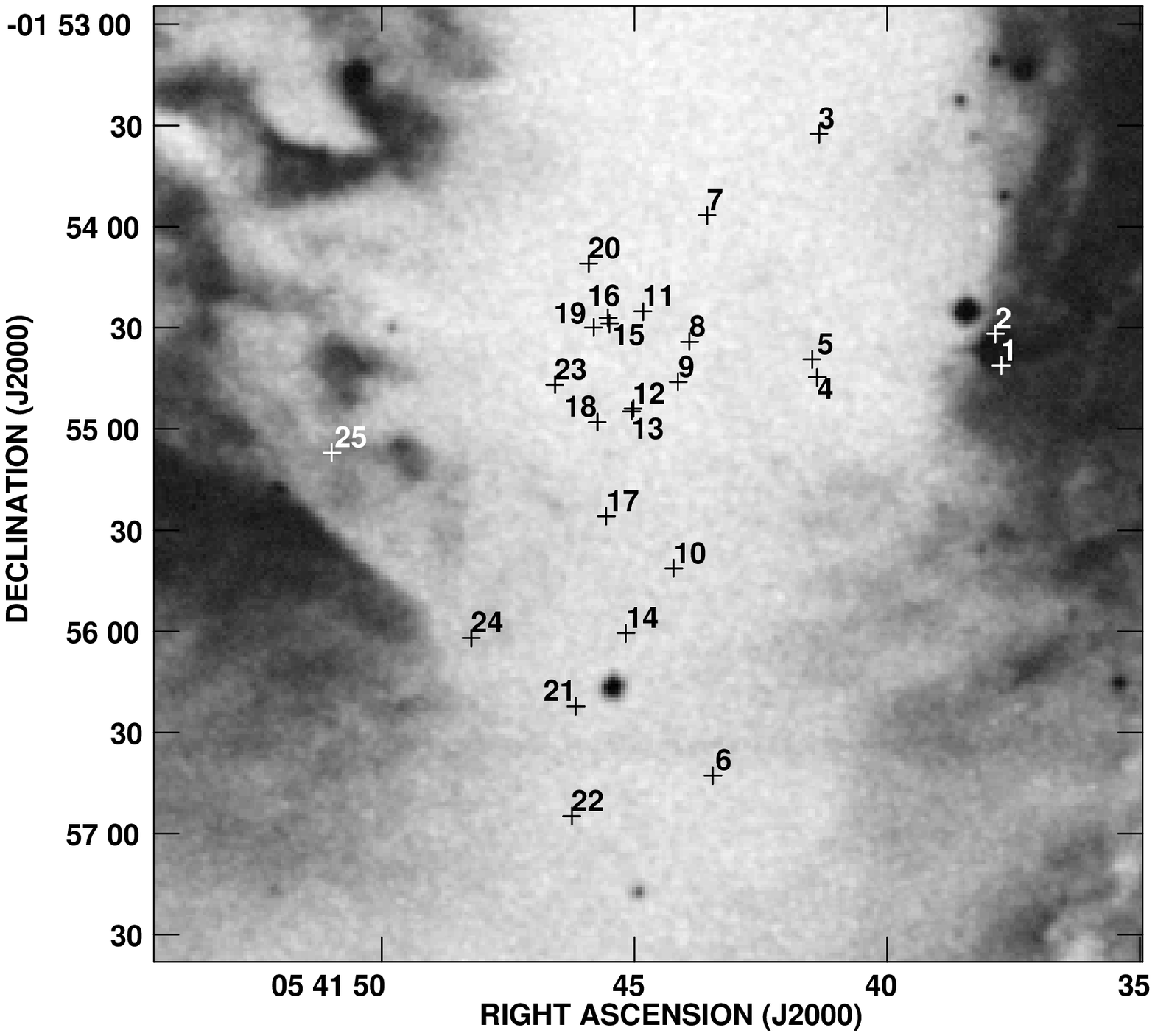}
\newpage
\plotone{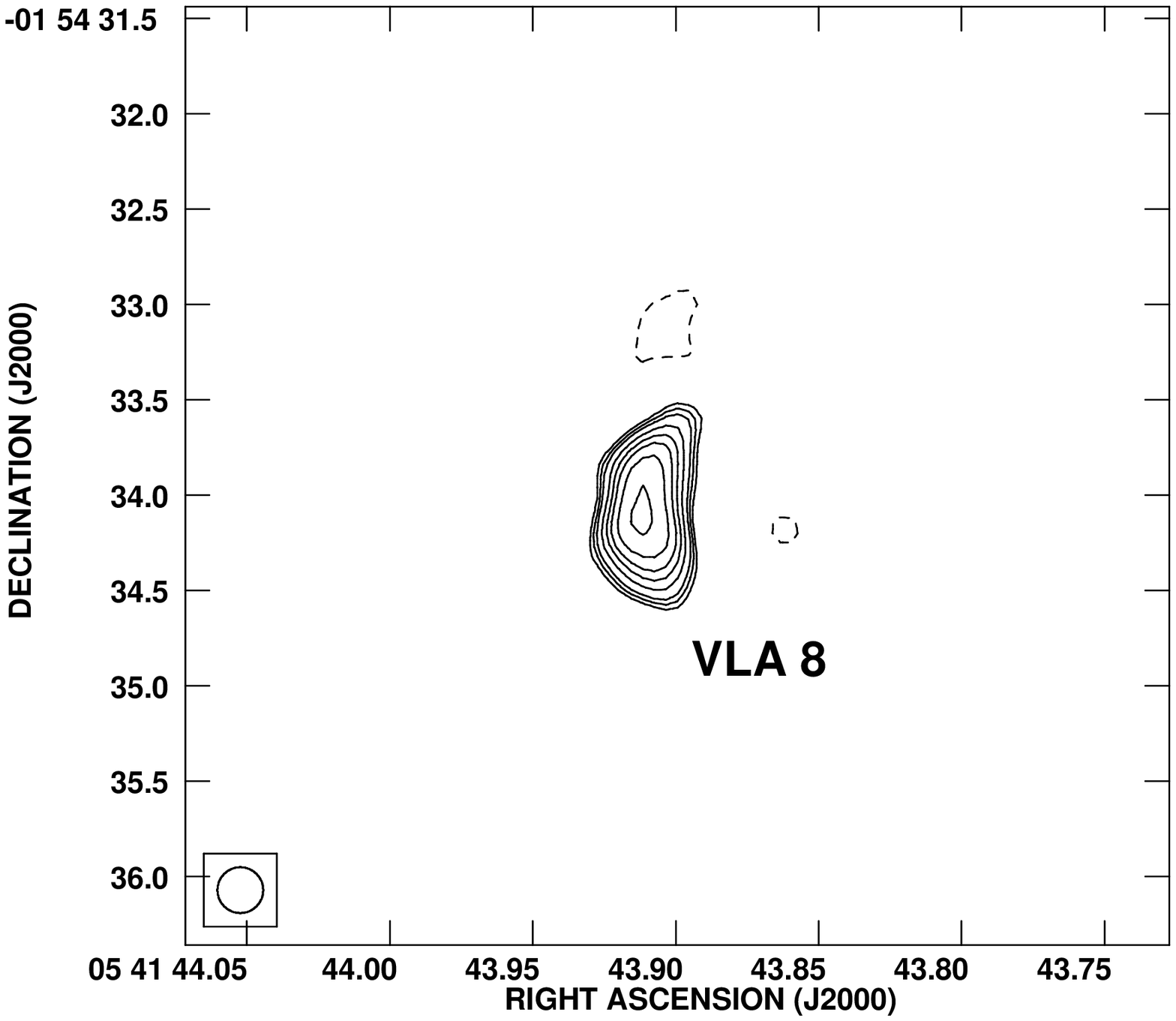}
\newpage
\plotone{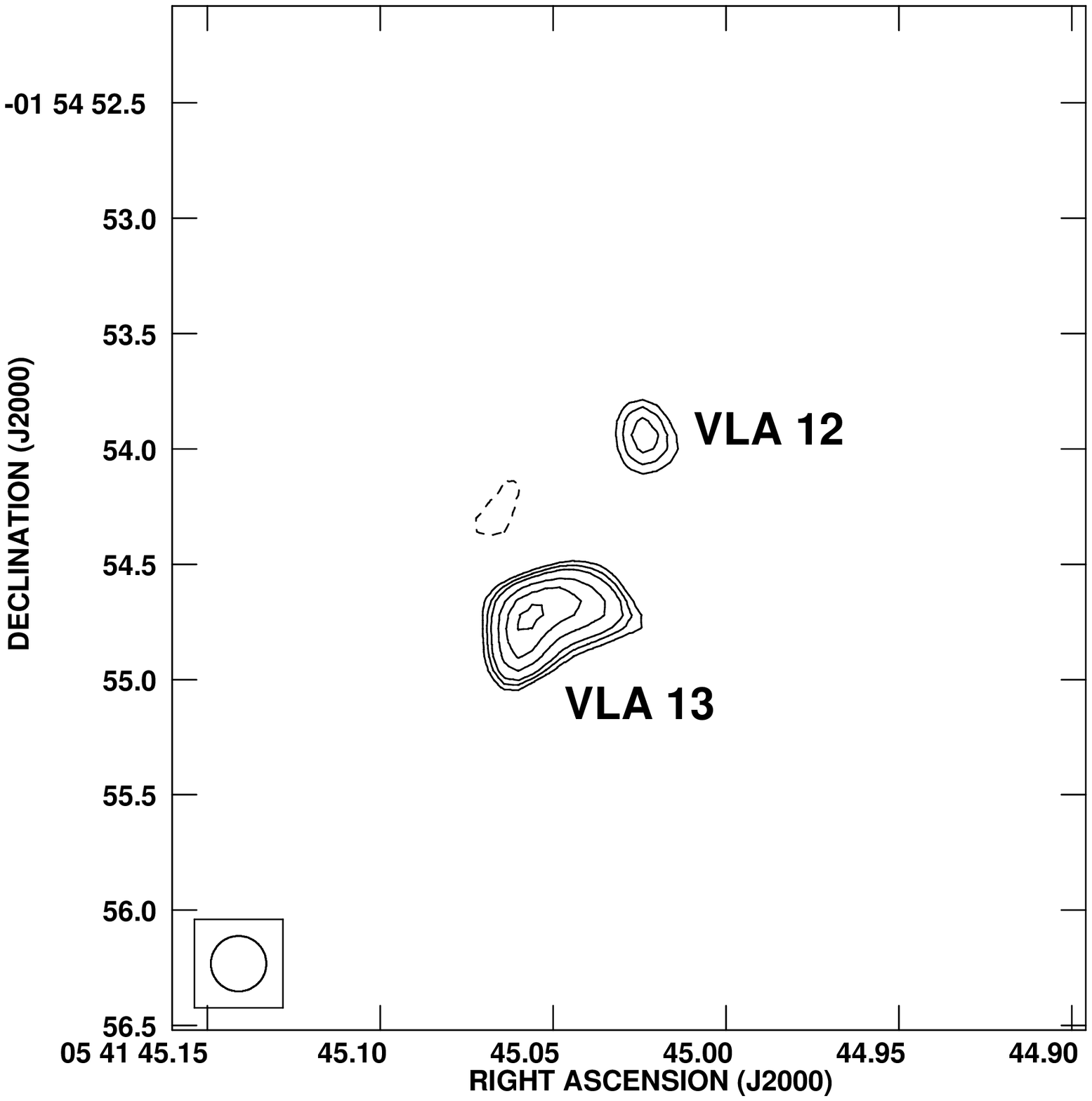}
\newpage
\plotone{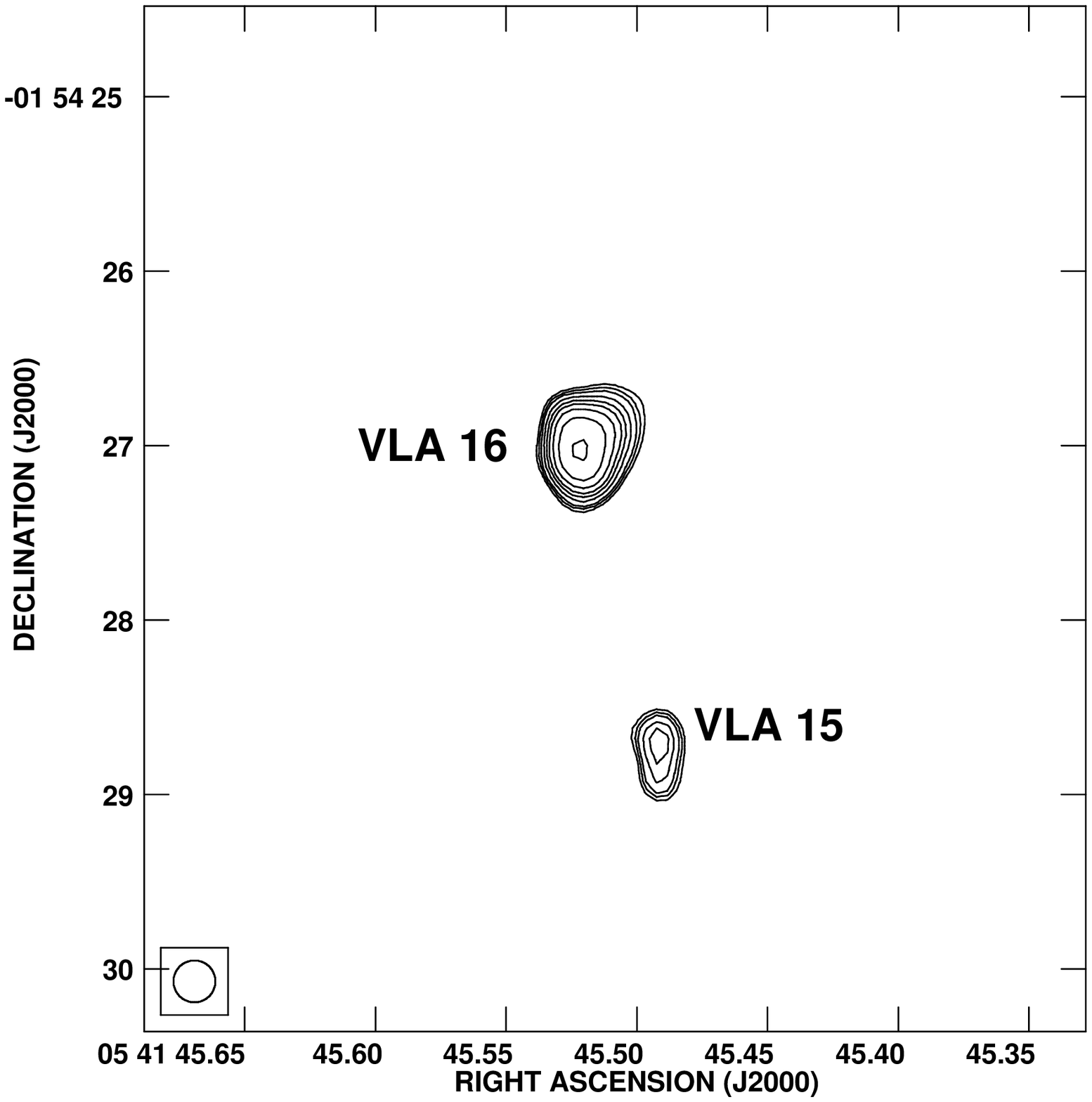}
\newpage
\plotone{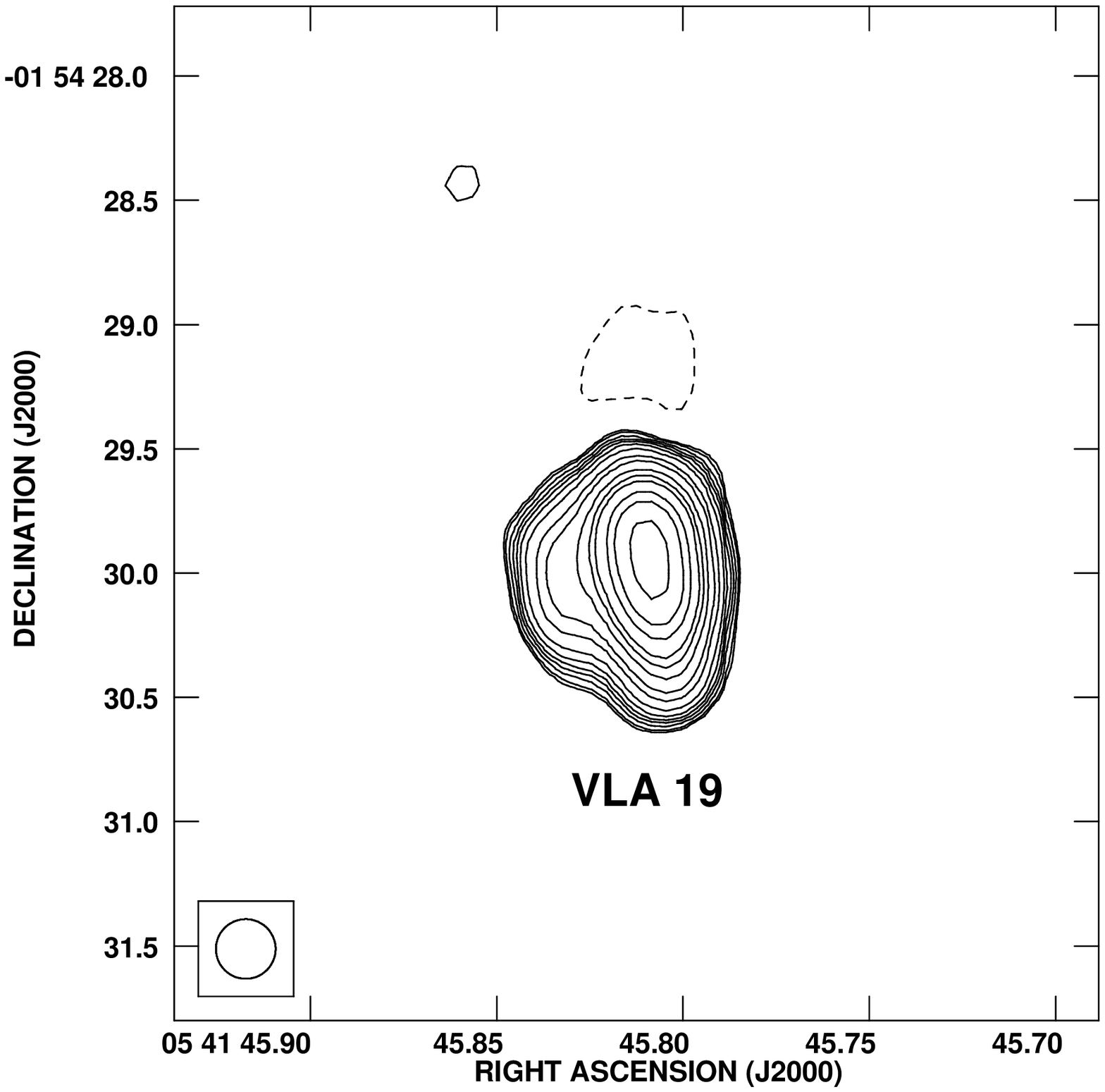}
\newpage
\plotone{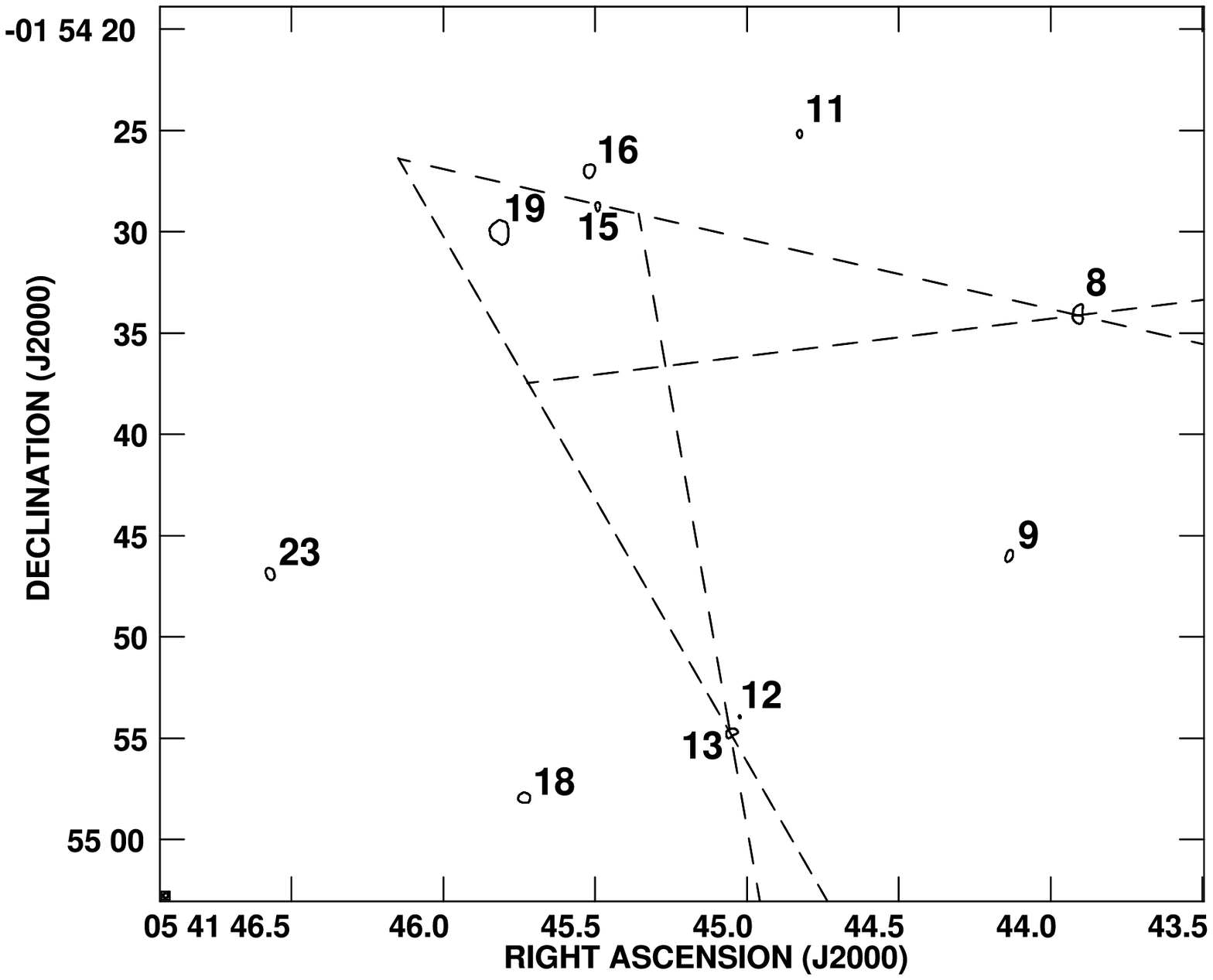}

\clearpage

\begin{deluxetable}{l c c c c c c c}
\tabletypesize{\scriptsize}
\tablecolumns{8} 
\tablewidth{0pc} 
\tablecaption{Parameters of the 3.6 cm VLA Sources}
\tablehead{
\colhead{}                              &                  
\colhead{R.A.}                          &
\colhead{Dec.}                          &
\colhead{Flux$^a$}                      &
\colhead{Time}                          &
\multicolumn{3}{c}{Counterpart$^b$}     \\
\cline{6-8}    
\colhead{VLA}                           & 
\colhead{$\alpha_{2000}$}               &
\colhead{$\delta_{2000}$}               &
\colhead{(mJy)}                         &
\colhead{Variable?}                     &
\colhead{2MASS}                         &
\colhead{Chandra}                       & 
\colhead{Other}                         \\
}
\startdata
 1  & 05 41 37.745 & -01 54 41.20 & 0.30 & N~~~~~ & & & HLP 2 \\
 2  & 05 41 37.858 & -01 54 31.82 & 0.16 & Y(1.9) & 05413786-0154323 & SGB 78 & \\
 3  & 05 41 41.346 & -01 53 32.54 & 0.28 & N~~~~~ & 05414134-0153326 & SGB 120 & HLL 52, BSC 69   \\
 4  & 05 41 41.385 & -01 54 44.64 & 0.26 & N~~~~~ & 05414138-0154445 & SGB 123 &   \\
 5  & 05 41 41.487 & -01 54 39.35 & 0.33 & Y(3.7) & 05414148-0154390 & SGB 124 & HLL 87, BSC 65   \\
 6  & 05 41 43.446 & -01 56 42.74 & 0.10 & Y(6.5) & 05414344-0156425 & SGB 147 & BCB IRS24   \\
 7  & 05 41 43.559 & -01 53 56.67 & 0.20 & N~~~~~ & 05414356-0153567 & SGB 152 &  \\
 8  & 05 41 43.912 & -01 54 34.13 & 0.63 & N~~~~~ & & &   \\
 9  & 05 41 44.136 & -01 54 46.04 & 0.33 & N~~~~~ & & & M FIR 4, MC89-4  \\
10  & 05 41 44.221 & -01 55 41.32 & 0.11 & N~~~~~ & & & M FIR 5, LCGR FIR 5 4   \\
11  & 05 41 44.827 & -01 54 25.16 & 0.22 & Y(1.6) & 05414482-0154251 & SGB 171 & HLL 77, HLP 22   \\
12  & 05 41 45.024 & -01 54 53.94 & 0.11 & N~~~~~~& & &   \\
13  & 05 41 45.056 & -01 54 54.73 & 0.24 & N~~~~~ & 05414504-0154546 & &   \\
14  & 05 41 45.168 & -01 56 00.56 & 0.29 & N~~~~~ & & & M FIR 6, LCGR FIR 6 n \\
15  & 05 41 45.492 & -01 54 28.70 & 0.24 & N~~~~~ & 05414550-0154286 & SGB 182 & B IRS2b \\
16  & 05 41 45.522 & -01 54 27.02 & 0.97 & N~~~~~ &  & SGB 183 &   \\
17  & 05 41 45.554 & -01 55 25.86 & 0.27 & N~~~~~ & & &   \\
18  & 05 41 45.733 & -01 54 57.93 & 0.49 & N~~~~~ & & &   \\
19  & 05 41 45.809 & -01 54 29.92 & 17.4 & N~~~~~ & 05414580-0154297 & SGB 187 & IRS2, KCW 206.543-16.347 \\
20  & 05 41 45.905 & -01 54 10.98 & 0.18 & Y(5.0) & & SGB 188 & BSC 93  \\
21  & 05 41 46.157 & -01 56 22.20 & 4.54 & Y(5.7) & & SGB 193 &   \\
22  & 05 41 46.236 & -01 56 54.80 & 0.10 & N~~~~~ & & SGB 196 &   \\
23  & 05 41 46.570 & -01 54 46.88 & 1.17 & Y(6.2) & 05414655-0154469 & SGB 200 &  \\
24  & 05 41 48.224 & -01 56 02.01 & 8.60 & Y(2.2) & 05414821-0156020 & SGB 210 &  \\
25  & 05 41 50.983 & -01 55 06.97 & 0.24 & N~~~~~ & 05415096-0155070 & & HLP 29, HLL 98 \\

\tablecomments{
                (a): Total flux density corrected for primary beam response.
The flux density reported is the average of the three epochs observed.\\
                (b): B = Bik et al. 2003;
                BCB = Barnes et al. 1989; BSC = Beck et al. 2003; 
                HLL = Haisch et al. 2000; HLP = Haisch et al. 2001;
                KCW = Kurtz et al. 1994; LCGR = Lai et al. 2002;
                M = Mezger et al. 1988;
                MC = Moore \& Chandler 1989;
                SGB = Skinner, Gagn\'{e}, \& Belzer 2003.\\ 
}
\enddata
\end{deluxetable}

\vfill\eject

\begin{deluxetable}{ l c c c c c }
\tabletypesize{\small}
\tablecolumns{6} 
\tablewidth{0pc} 
\tablecaption{Parameters of Radio Clusters in Regions of Recent Star Formation}
\tablehead{
\colhead{}                        & 
\colhead{Luminosity}         &
\colhead{Number of}                & 
\colhead{Most Luminous}        &
\colhead{Diameter}       &
\colhead{} \\ 
\colhead{Region}       &
\colhead{($L_\odot$)}       &
\colhead{Members}        &
\colhead{Member (mJy~kpc$^2$)$^a$}       &
\colhead{(pc)}       &
\colhead{Reference}        \\
}
\startdata
Orion & $\sim 2 \times 10^5$  & $\sim$50 & 8.8 & 0.3 & Menten \& Reid (1995)  \\
NGC~1579 & $\sim 2 \times 10^3$ & 16 & 0.6 & 0.3 & Stine \& O'Neal (1998) \\
GGD~14 & $\sim 10^4$ & 6 & 0.2 & 0.2 & G\'omez et al. (2002)  \\
NGC~1333 & $\sim 120$ & 44 & 0.3 & 0.7 & Rodr\'\i guez et al. (1999) \\
Arches & $\sim 10^7$ & 8 & 122.8 & 0.7 & Lang et al. (2001) \\
NGC 2024 & $\sim 5 \times 10^4$  & 25 & 3.0 & 0.5 & This paper \\

\tablecomments{
                (a): 3.6 cm flux density times distance in kpc squared \\
}
\enddata
\end{deluxetable}

\vfill\eject

\clearpage

\end{document}